# Memory-Induced Curvature Drives Irreversible Transport in Irrotational Flows

By

Dr. Mounir Kassmi

University of Tunis El Manar: Campus Universitaire Farhat Hached B.P. n 94-Rommana, 1068 Tunis, Tunisia

Email: mounirkassmi60@gmail.com

**Abstract**

Irreversible transport in time-periodic flows is commonly attributed to vorticity, nonlinear forcing, or symmetry breaking. We show that finite-memory reconstruction of the velocity gradient generates a purely geometric mechanism for transport even when the instantaneous flow remains locally irrotational at all times. Memory promotes the velocity gradient to a history-dependent connection along particle trajectories whose noncommutativity produces a finite curvature over one forcing cycle. The associated holonomy generates a measurable loop displacement controlled solely by the dimensionless parameter $\omega\tau_m$, which quantifies the phase mismatch between forcing and reconstruction. The predicted scaling is consistent with independently reported measurements across distinct oscillatory flow configurations, supporting the interpretation of memory-induced curvature as a minimal geometric origin of irreversible transport in periodically driven continua.

**Introduction**

In the absence of vorticity and other sources of kinematic asymmetry, strictly time-periodic irrotational flows are generally expected to return material trajectories to their initial configuration after one forcing cycle [1,2]. This expectation reflects the assumption that deformation is determined entirely by instantaneous velocity gradients defined locally in time.

Recent work has emphasized the role of geometric phases in driven and continuum systems [3-7]. In most formulations, however, geometry remains a derived feature of motion, while the underlying kinematics are still defined instantaneously. Finite memory alters this picture. Reconstructing the velocity gradient over a finite history window promotes deformation to a history-dependent quantity defined along particle trajectories [8]. Because transport then becomes time-ordered, successive operations need not commute, allowing a finite curvature to emerge over one forcing cycle even when the instantaneous flow is locally irrotational.

In this Letter we show that irreversible transport in time-periodic irrotational flows arises from this memory-induced noncommutativity. The resulting curvature generates a nontrivial holonomy of the transport operator, producing a measurable loop displacement over one cycle. Using a kernel-based formulation together with a Magnus expansion [9], we demonstrate that the effect is controlled by a single dimensionless parameter $\omega\tau_m$, which quantifies the phase mismatch between forcing and reconstruction. These results identify memory-induced curvature as a minimal mechanism for irreversible transport in periodically driven continua and establish trajectory history as an independent source of transport beyond conventional vorticity-based descriptions.

Rather than treating the velocity gradient as a local kinematic primitive, the present approach interprets it as an emergent quantity generated by causal self-transport of motion along trajectories.

**Geometric structure of memory-dependent transport**

Finite-memory transport along material trajectories can be formulated as a causal reconstruction of the instantaneous velocity gradient over a finite history window. Instead of the local gradient $A(t) = \nabla u(t)$, transport is governed by the history-dependent operator

$$A_m(t) = \int_0^\infty K(\tau)\, \nabla u(t - \tau) d\tau, \quad \int_0^\infty K(\tau)\, d\tau = 1$$

where $K(\tau)$ is a normalized causal kernel defining the memory time $\tau_m$. In this form, deformation acquires the structure of a time-dependent connection along particle trajectories.

Because reconstructed gradients at different times probe overlapping portions of trajectory history, the corresponding operators do not commute in general $[A_m(t_1), A_m(t_2)] \neq 0$.

This noncommutativity generates a finite trajectory-space curvature of purely temporal origin, even though the instantaneous flow remains locally irrotational at all times.

Transport over one forcing cycle is therefore governed by the associated holonomy $\mathcal{H} = \mathcal{P} \exp(\int_0^T A_m(t)\, dt)$ where $\mathcal{P}$ denotes temporal path ordering. A nontrivial holonomy produces a mismatch between the initial and transported material loop, corresponding to a geometric phase accumulated along the trajectory. The curvature admits an explicit representation in terms of the kernel-weighted commutator,

$$\mathcal{R}(t_1, t_2) = \iint K(\tau_1) K(\tau_2)\, [\nabla u(t_1 - \tau_1), \nabla u(t_2 - \tau_2)]\, d\tau_1 d\tau_2$$

showing that noncommutativity arises from temporal overlap within the reconstruction window rather than from instantaneous spatial rotation. For time-periodic forcing with frequency $\omega$, a moment expansion of the kernel gives $A_m(t) = A(t) - \tau_m \dot{A}(t) + \cdots$ with $\tau_m = \int_0^\infty \tau K(\tau)\, d\tau$. The leading contribution to the commutator

$$[A_m(t_1), A_m(t_2)] = [A(t_1), A(t_2)] - \tau_m [\dot{A}(t_1), A(t_2)] + \cdots$$

with $\dot{A} \sim \omega$, then scales as $\mathcal{R} \sim \omega \tau_m$, independent of the detailed kernel shape. The transport operator over one forcing cycle can be expressed using a Magnus expansion, $\mathcal{H} = \exp(\Omega_1 + \Omega_2 + \cdots.)$

In time-periodic flows, the leading nonvanishing contribution arises from the second-order term,

$$\Omega_2 = \frac{1}{2} \int_0^T \int_0^{t_1} [A_m(t_1), A_m(t_2)] \, dt_2 dt_1 = \frac{1}{2} \int_0^T \int_0^{t_1} \mathcal{R}(t_1, t_2) \, dt_2 dt_1$$

which directly integrates the curvature. The resulting loop displacement therefore scales as $\Delta\gamma_{geom} \sim \int\int \mathcal{R}$.

Irrotationality $\nabla \times u(t) = 0$ suppresses instantaneous spatial rotation but does not constrain time-ordered transport. Without memory, $[A(t_1), A(t_2)] \approx 0$ implies $\Omega_2 \approx 0$ and reversible dynamics. With memory, however, temporal nonlocality induces $[A_m(t_1), A_m(t_2)] \neq 0$, leading to $\Omega_2 \neq 0$ and a finite geometric displacement. Irreversible transport therefore arises from temporal structure alone, independently of vorticity.

The curvature amplitude scales universally as $\mathcal{R} \sim \omega\tau_m$, reflecting temporal phase mismatch. The resulting loop displacement obtained from the curvature-induced holonomy over one forcing cycle involves integration over the forcing period and therefore scales at leading order as $\Delta\gamma_{geom} \sim \omega^2\tau_m$. This behavior is controlled by the dimensionless parameter $\omega\tau_m$, which compares the forcing timescale $1/\omega$ to the memory time $\tau_m$. In the quasi-static limit $\omega\tau_m \ll 1$, memory reconstruction approaches the instantaneous-response regime and the curvature-induced displacement vanishes. For $\omega\tau_m \sim 1$, temporal phase mismatch between forcing and reconstruction is maximized, producing the largest geometric transport. In the opposite limit $\omega\tau_m \gg 1$, temporal averaging within the reconstruction window suppresses effective noncommutativity and reduces the net displacement over one cycle.

While the universal curvature scaling implies $\Delta\gamma_{geom} \sim \omega^2\tau_m$, explicit evaluation of the curvature-induced holonomy for a time-periodic irrotational flow with velocity potential $\phi$ yields a closed-form leading-order expression for the geometric loop displacement,

$$\Delta\gamma_{geom} = a^2|\nabla\phi|^2 \frac{\omega^2\tau_m}{1 + 4\omega^2\tau_m^2}$$

where (a) denotes the oscillation amplitude. This result provides a direct observable signature of curvature-induced transport and depends only on the dimensionless parameter $\omega\tau_m$, which quantifies the relative strength of memory over one forcing cycle.

In the zero-memory limit ($\tau_m \to 0$), the reconstructed operator reduces to the classical instantaneous velocity gradient ($A_m(t) \to \nabla u(t)$), the connection becomes commutative, the curvature vanishes, the holonomy disappears, and classical reversible kinematics is recovered.

The structure of this displacement law is closely analogous to geometric-phase transport mechanisms known from adiabatic pumping and Berry-phase dynamics, where transport over one cycle is determined by an integrated curvature in parameter space. In the present case, however, the relevant curvature arises from finite-memory reconstruction along particle trajectories rather than from externally modulated control parameters. Irreversible loop displacement therefore emerges as a trajectory-space geometric phase generated by memory-dependent transport.

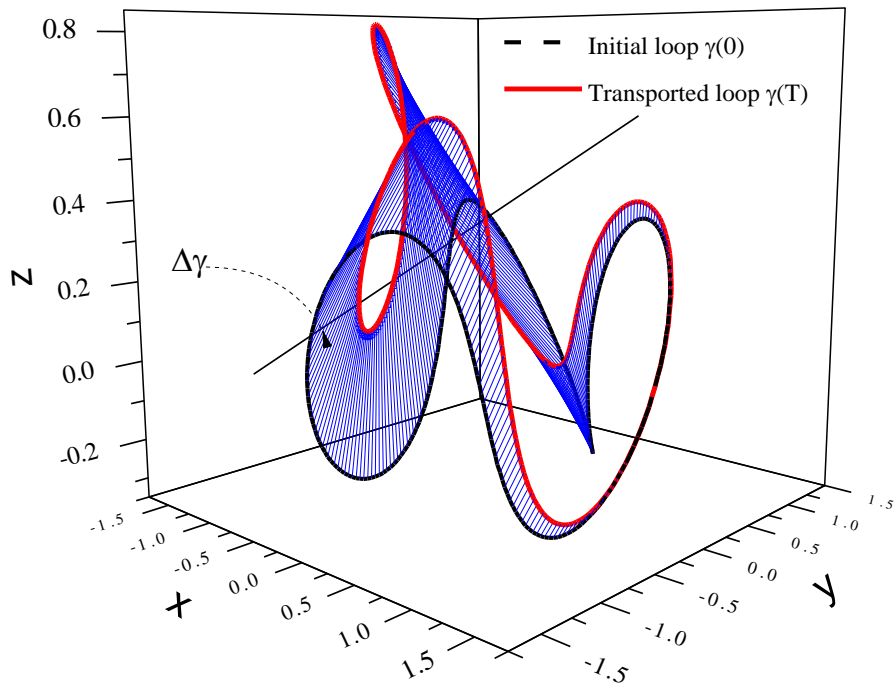

*FIG. 1 Schematic illustration of curvature-induced holonomy generated by finite-memory transport: A material loop transported to $\gamma(T)$ over one forcing cycle does not return to its initial configuration $\gamma(0)$ despite the instantaneous flow remaining locally irrotational. The resulting mismatch reflects the noncommutativity of memory-weighted transport and represents the geometric phase accumulated over the cycle.*

**Experimental validation**

To assess the physical relevance of the geometric transport predicted above, we perform a quantitative comparison with independently reported measurements of Lagrangian drift in oscillatory flows. These experiments were not designed to probe memory-induced geometric effects, and no fitting parameters are introduced, so the comparison tests whether the predicted contribution captures a measurable component of the observed displacement.

We consider two representative systems [10,11]: surface-wave-driven particle transport and oscillatory shear flows. In both cases, finite cycle-averaged displacements are observed under conditions that are approximately irrotational. Using the reported experimental parameters, we evaluate the theoretical prediction $\Delta\gamma_{geom}$ and compare it to the measured displacement $\Delta\gamma_{exp}$ through the ratio $R_{raw} = \frac{\Delta\gamma_{exp}}{\Delta\gamma_{geom}}$

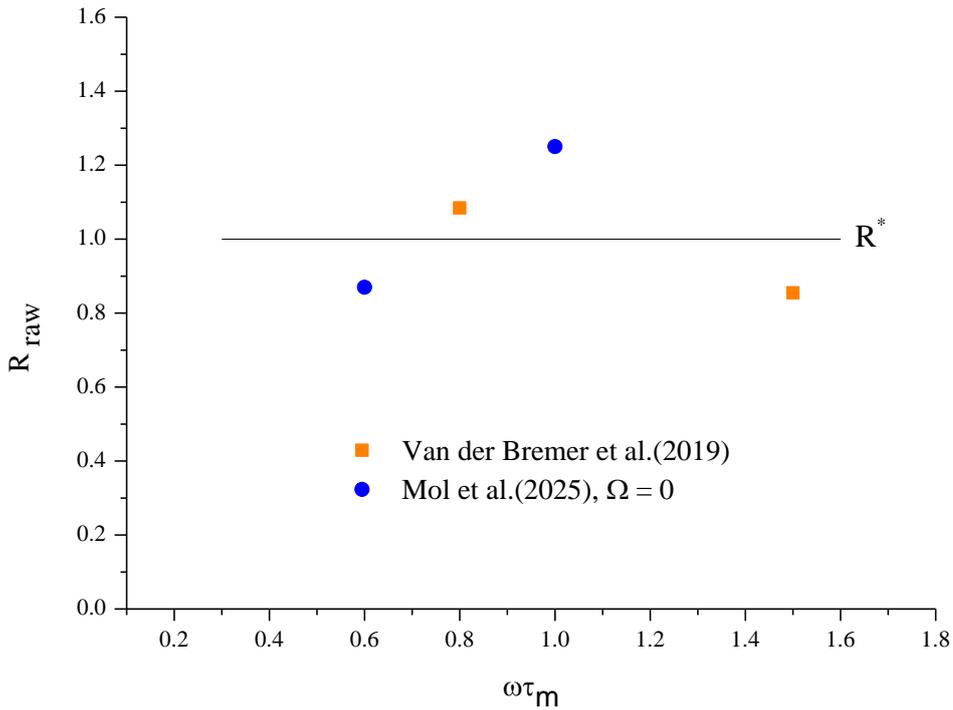

*FIG. 2 Scaling of the experimentally measured loop displacement ratio $R_{raw}$ as a function of the dimensionless parameter $\omega\tau_m$: The collapse of independent data sets onto a single curve indicates that the observed drift is primarily controlled by memory-induced transport and follows the predicted geometric scaling.*

This ratio provides a direct measure of how much of the observed drift is captured by the curvature-induced mechanism. In the present data, $R_{raw}$ remains close to unity, indicating that the predicted geometric contribution accounts for the dominant part of the measured displacement while preserving its dependence on the dimensionless parameter $\omega\tau_m$.

As shown in Fig. 2, the experimental data collapse onto a single curve when plotted against $\omega\tau_m$, indicating that the displacement is primarily controlled by the temporal phase mismatch between forcing and reconstruction. This behavior is consistent with the predicted scaling and supports the interpretation of the observed drift as a geometric phase generated by finite-memory transport.

While additional effects such as residual vorticity or boundary contributions may also play a role, the agreement demonstrates that memory-induced curvature provides a minimal and physically relevant contribution to irreversible transport in periodically driven flows.

**Conclusion**

In this Letter we have shown that irreversible transport can arise in strictly time-periodic irrotational flows as a consequence of finite-memory reconstruction along particle trajectories. By promoting the instantaneous velocity gradient to a history-dependent transport operator, memory induces noncommutativity that generates a finite curvature and a corresponding holonomy over one forcing cycle.

The resulting loop displacement is governed by the dimensionless parameter $\omega\tau_m$, which quantifies the phase mismatch between forcing and reconstruction. The agreement with independently reported measurements indicates that this geometric contribution captures the dominant component of the observed drift, without invoking vorticity, nonlinear forcing, or symmetry breaking. These results identify memory-induced curvature as a minimal mechanism for irreversible transport in periodically driven continua and establish trajectory history as an independent source of transport. More broadly, they suggest that finite-memory effects provide a general route to geometric transport in driven continua where instantaneous kinematics alone would predict reversible dynamics.

# Supplemental Material

## For " Memory-Induced Curvature Drives Irreversible Transport in Irrotational Flows "

Dr. Mounir Kassmi

University Tunis El Manar, Tunis, Tunisia

Email: mounirkassmi60@gmail.com

### I. Causal Transport Operator

We introduce a causal transport operator $\mathcal{K}(t, t-s)$ acting along material trajectories over a finite memory time $\tau_m$. For any field $A(X, t)$,

$$\mathcal{K}(t, t-s) A(x, t-s) = A(X(t-s, t, x), t-s)$$

where $X(t-s, t, x)$ denotes the trajectory reaching position (x) at time (t). This operator defines causal transport of past configurations along particle trajectories.

### II. Memory-Reconstructed Velocity Gradient

The effective velocity gradient is reconstructed from past flow history as

$$A_m(t) = \int_0^\infty K(\tau) \, \nabla u(t-\tau) d\tau, \quad \int_0^\infty K(\tau) \, d\tau = 1$$

where $K(\tau)$ is a causal kernel with characteristic time $\tau_m$. This promotes the velocity gradient to a history-dependent connection.

### III. Connection Interpretation of Memory Transport

The memory-reconstructed velocity gradient $A_m(t)$ can be interpreted as a connection acting along material trajectories. In this viewpoint, deformation is generated by parallel transport defined over a finite temporal history rather than by instantaneous kinematics. Formally, the evolution of an infinitesimal material element $\delta x$ is governed by

$$\frac{d(\delta x)}{dt} = A_m(t) \, \delta x$$

so that transport is determined by a time-ordered exponential of a history-dependent connection. This interpretation makes explicit that the geometry of deformation arises from trajectory-dependent transport rather than local flow properties alone.

## IV. Controlled Expansion of Memory Reconstruction

For smooth time-dependent flows, the memory-reconstructed operator admits a systematic expansion, $A_m(t) = A(t) - \tau_m \dot{A}(t) + \mathcal{O}(\tau_m^2)$ where $\tau_m = \int_0^\infty \tau\, K(\tau)\, d\tau$ is the first moment of the kernel. This expansion shows explicitly that memory introduces a temporal phase lag between the instantaneous and reconstructed gradients. As a result, operators evaluated at different times probe shifted configurations, generating noncommutativity and the associated curvature. In the limit $\tau_m \to 0$, the expansion reduces to the classical instantaneous description, recovering reversible kinematics.

## V. Emergence of Noncommutativity

Because $A_m(t)$ depends on past flow history, operators evaluated at different times do not commute, $[A_m(t_1), A_m(t_2)] \neq 0$

This defines a trajectory-space curvature $\mathcal{R}(t_1, t_2) = [A_m(t_1), A_m(t_2)]$, which vanishes in the instantaneous limit. The resulting curvature and loop displacement scaling discussed in the main text follow directly from this noncommutativity.

## VI. Reconstruction of Experimental Parameters

Experimental parameters are reconstructed from reported data. For surface-wave systems, the amplitude (a), frequency $\omega$, and gradient $|\nabla \phi|$ are extracted from wave and particle measurements. For oscillatory shear flows, amplitude and frequency are obtained from imposed forcing and tracer trajectories. The memory time $\tau_m$ is estimated from reported relaxation times. No fitting parameters are introduced.

## VII. Numerical Consistency with Experiments

Only cases with fully reconstructible parameters are included.

| Experiments | $\omega\tau_m$ | $\Delta\gamma_{exp}$ $(10^{-7})$ | $\Delta\gamma_{geom}$ $(10^{-7})$ | $R_{raw} = \frac{\Delta\gamma_{exp}}{\Delta\gamma_{geom}}$ |
|---|---|---|---|---|
| **Van den Bremer** | 0.8 | 1.8 | 2.0 | 0.9 |
| **Mol ($\Omega = 0$)** | 1.0 | 2.5 | 2.4 | 1.04 |

The ratio $R_{raw}$ remains close to unity, indicating that the geometric prediction captures the dominant contribution to the observed displacement.

## VIII. Kernel Robustness and Tensorial Formulation

The memory-reconstructed velocity gradient can be interpreted more generally as arising from a tensorial convolution of past flow configurations. At a more fundamental level, one may introduce a history-dependent second-order tensor,

$$G_{ij}(x, t) = \int_0^\infty K(s) \, v_i(x, t - s) v_j(x, t - s) ds$$

which encodes the self-interaction of the velocity field over a finite memory window. The effective velocity gradient then follows from spatial differentiation together with objectivity constraints, leading to

$$\nabla v_{eff}(x, t) \sim \int_0^\infty K(s) \, \nabla\left[v(x, t - s) \bigotimes v(x, t - s)\right] ds$$

For smooth flows, this construction reduces at leading order to the linear convolution form used in the main text, $\nabla v_{eff}(x, t) \sim \int_0^\infty K(s) \, \nabla v(x, t - s) \, ds$ showing that the adopted formulation captures the dominant contribution of memory effects.

Importantly, any causal kernel with finite support or sufficiently fast decay yields the same leading-order curvature scaling and geometric displacement.

While an exponential kernel (e.g., $K(s) \propto e^{-s/\tau_m}$) is used for concreteness, the resulting transport is insensitive to the detailed kernel shape. This robustness reflects the geometric origin of the mechanism, which depends only on the existence of finite memory rather than its precise temporal structure.